\documentclass[aps,showpacs,showkeys]{revtex4}
\usepackage{graphicx}
\begin{document}



\title{ A discrete inhomogeneous model for the yeast cell cycle}

\author{Lucas Wardil$^{\dag\dagger}$, and
Jafferson Kamphorst L. da Silva$^{\dag\ast}$}
\affiliation{$^\dag$Departamento de F\'\i sica, Universidade Federal de Minas Gerais, \\
Caixa Postal 702, CEP 30161-970, Belo Horizonte - MG, Brazil\\}
\date{\today}




\begin{abstract}

 We study the robustness and stability of the yeast cell regulatory network  by using a general 
inhomogeneous discrete model. We find that inhomogeneity, 
on average, enhances the stability of the biggest attractor of the dynamics and that the large size of 
the basin of attraction is robust against changes in the parameters of inhomogeneity. 
We  find that the most frequent orbit, which represents the cell-cycle pathway, has a better biological 
meaning than the one exhibited by the homogeneous  model. 
\end{abstract}

\maketitle
\section{INTRODUCTION}

The eukaryotic cell exhibits a common process of division into two daugther cells.
 This process consists of four phases \cite{alberts}:
(i) G1 phase, in which the cell grows;
(ii) S phase, in which the DNA is replicated; (iii) G2 phase, that is a temporally gap between the S phase
 and the next one; (iv) M phase, in which the cell divides itself in two cells. In G1 phase the cell cycle 
 rests in a stationary state until the cell size reachs a critical value and the cell receive external
  signals which allow it to go on the cycle.
 The cell-cycle regulation machinery, which controls the growth and division processes,
 is  known for the budding yeast in detail, {\sl Saccharomyces cerevisiae} \cite{chen,li}.
 In order to understand the budding yeast regulation, several models have been proposed and
 discussed \cite{chen,li,cross,chen2,chen3,cross2,futcher,murray,ingolia,tyers}.
 Chen et al. \cite{chen} converted the regulation mechanism into a set of differential equations
 with empirical parameters. Their kinetic model has taken in account  several physiological,
  biochemistry and genetical details.
  Recently, Li et al. \cite{li}  introduced a simple Boolean dynamical model to investigate
 the stability and robustness features of the regulatory  network. They found that the cell-cycle network
 is stable and robust.  A stochastic version of this model, controlled by a temperature like
 parameter,  was latter studied \cite{zhang}. The authors found that the stationary state and the
 cell-cycle pathway are stable for a wide range of the temperature parameter values.
 Other aspects related to the checkpoint efficiency of the Li et al. model were also
 considered \cite{stoll}.

In this work we consider the inhomogeneous version of the yeast cell-cycle introduced
 by Li et al. \cite{li}. The cell cycle is represented by a regulatory network and the dynamics is modeled 
 as a simple discrete dynamical system. The dynamics is constrained by the network topology by means of some 
 parameters (coefficients) related to each network link. The link between nodes $i$ and $j$
determines the value of parameter $K_{i,j}$, which is present in the time evolution rule. 
 If the coefficient is positive, the link represents an activation and it will be noted as $K_{i,j}=a(i,j)$
  ($a(i,j)>0$).
 On the other hand, a negative coefficient $K_{i,j}=-b(i,j)$ ($b(i,j)>0$) represents an inactivation link.
In the paper of Li et al. \cite{li}, the authors studied basically the
homogeneous model $a(i,j)=b(i,j)=1$ for the nodes with a link between them. At most they have considered only
two kinds of intensity $a(i,j)=a_r$ and $b(i,j)=a_g$.  However, the interactions are important for
 the dynamics and are related to the constant rates of the kinetic equations, implying that
 different  contributions to activation or inactivation may be present. We consider the most general 
  inhomogeneous model of the yeast cell-cycle. 
 Since that several different inhomogeneities represent the same dynamical model,
 we eliminate all possible degeneracy by constructing a minimal set of parameters (coefficients).
 Such set represents all kind of inhomogeneity and is very large. We find that the  big basin of 
 attraction corresponding to the stationary state is still robust to changes in the coefficients and 
 that in the minimal set of coefficients a new orbit corresponding to the cell cycle  appears more 
 frequently having a more feasible biological significance. This means that this orbit is more robust
  against change in coefficients than the orbit of the homogeneous model. Moreover,
   the mean basin size of the global attractor is bigger than
   for the homogeneous case, when this more frequent orbit is present.

\section{THE DYNAMICAL MODEL}

\begin{figure}
  \centering
  \scalebox{.3}{\includegraphics{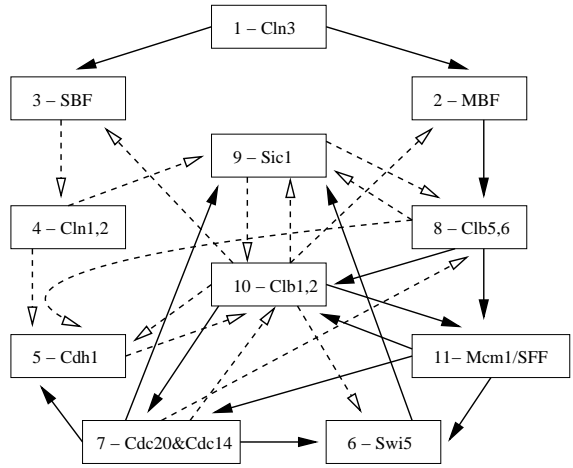}}
\caption{The yeast cell cycle regulatory network\cite{li}. The nodes are identified by the proteins or complex 
proteins and by the number used in the definition of the dynamical system. The continuous arrows mean 
activation and the dashed ones mean inhibition.}\label{network}
\end{figure}

Our model is based on the deterministic Boolean  model of Li et al. \cite{li} for
the budding yeast cell-cycle regulation, which is represented by a regulatory network \ref{network}. 
Each one of the 11 nodes, which represents
proteins or protein complexes,  is represented by a variable $S$ that takes the values
$0$ (the protein state is inactive) or $1$ (the protein state is active). The configuration of the system at
time $t$ is described by a vector ${\vec S}(t)=(S_1(t),S_2(t),\ldots,S_{11}(t))$ that
represents the state of the following proteins or complex proteins:  (Cln3, MBF, SBF, Cln1-2, Cdh1, Swi5,
 Cdc20, Clb5-6, Sic1, Clb1-2, Mcm1). A configuration can be expressed in a short fashion  by an integer 
 $I$ if we define that $I=\sum_{i=1}^{11}S_i2^{i-1}$. For example, if only Cdh1 and Sic1 are active
 ($S_5=1$ and $S_9=1$), the related
 configuration ${\vec S}=(0,0,0,0,1,0,0,0,1,0,0)$ corresponds to $I({\vec S})=272$.

The state vector  ${\vec S}$ describes the cell state in a specific time. If one wants to capture the 
time evolution of the cell states, one needs to address a dynamical model, in which there is a time 
evolution rule. The configuration at time $t+1$,
${\vec S}(t+1)$ is
related to the previous one ${\vec S}(t)$ by the relationships
\begin{equation}
 S_i(t+1)= \left\{     \begin{array}{ll}
                        1,     &\mbox{if $\sum_j K_{i,j} S_j(t) > 0$} \nonumber\\
                        0,     &\mbox{if $\sum_j K_{i,j} S_j(t) < 0$} \label{eq.dynamics}\\
                       S_i(t), &\mbox{if $\sum_j K_{i,j} S_j(t) = 0$}~~,\nonumber
                          \end{array}
                  \right.
\end{equation}
where $K_{i,j}$ is the interaction between nodes $i$ and $j$ and $i=1,2,\ldots ,11$.
If there is no link between the nodes,
 we have that $K_{i,j}=0$. Let us set our notation. When this link represents
an activation, $K_{i,j}=+a(i,j)$. Here $a(i,j)$ is the intensity of such activation.
When the interaction represents an inhibition we have that $K_{i,j}=-b(i,j)$, with $b(i,j)$ being
the inhibition intensity. Note that nodes $1,~4,~6,~7$ and $11$ have also a time-delayed self-degradation 
mechanism.
 When $S_i(t)=1$ ($i=1,4,6,7,11$) and $\sum_{j}K_{i,j}S_j=0$ from time $t$ to $t+t_d$,
 we will have that $S_i(t+t_d)=0$. From now on we will consider $t_d=1$. This evolution rule can be
  set in a more compact form:
\begin{equation}
\begin{array}{ccc}
S_i(t+1)=F_i\left(\sum_j K_{i,j} S_j(t)\right), & \mbox{where} & F_i(x)=\left\{     \begin{array}{ll}
                        1,     & \mbox{if $x > 0$}  \nonumber\\
                        0,     & \mbox{if $x < 0$}\label{eq.dynamics2}\\
                       S_i(t), & \mbox{if $x = 0$}~~.\nonumber
                          \end{array}
                  \right.
\end{array}
\end{equation}

If we name $\Omega$ as the entire state space vector, which is the space that contains the 
$2^{11}=2048$ possible vectors ${\vec S}(t)$, and if we name $F$ 
the map defined by $\vec{S}(t+1)=F(\vec{S}(t))=(F_1(S_1(t)),\ldots,F_{11}(S_{11}(t)))$, 
we can define a dynamical systems as the pair $(\Omega,F)$, em que $F: \Omega \rightarrow \Omega$.

From Fig. \ref{network} we can obtain the dynamical equations. For further use,
let us write these equations as the trivial  ones, namely
\begin{eqnarray}
S_1(t+1)&=& 0~~,\nonumber\\
S_4(t+1)&=& F_4(a(4,3)S_3(t))~~,\nonumber\\
S_{7}(t+1)&=& F_7(a(7,10)S_{10}(t)+a(7,11)S_{11}(t))~~,\label{eq.trivial}\\
S_{11}(t+1)&=& F_{11}(a(11,8)S_8(t)+a(11,10)S_{10}(t))~~\nonumber,
\end{eqnarray}
those involving one positive and one negative links
\begin{eqnarray}
S_2(t+1)&=& F_2(a(2,1)S_1(t)-b(2,10)S_{10}(t))~~,\label{eq.two-a}\\
S_3(t+1)&=& F_3(a(3,1)S_1(t)-b(3,10)S_{10}(t))~~\label{eq.two-b},
\end{eqnarray}
the ones involving one positive and two negative links, and the symmetrical case,
\begin{eqnarray}
S_6(t+1)&=& F_6(a(6,7)S_7(t)+a(6,11)S_{11}(t)-b(6,10)S_{10}(t))~~,\label{eq.three-a}\\
S_8(t+1)&=& F_8(a(8,2)S_2(t)-b(8,7)S_7(t)-b(8,9)S_9(t))~~\label{eq.three-b},
\end{eqnarray}
 one equation involving three negative and one positive links
\begin{equation}
S_5(t+1)=F_5(a(5,7)S_7(t)-b(5,4)S_4(t)-b(5,8)S_8(t)-b(5,10)S_{10}(t))~~\label{eq.four},
\end{equation}
and, finally, those involving three negative and two positive parameters
\begin{eqnarray}
S_9(t+1)&=& F_9(a(9,6)S_6(t)+a(9,7)S_{7}(t)-b(9,4)S_{4}(t)-b(9,8)S_{8}(t)-b(9,10)S_{10}(t))~~,
\label{eq.five-a}\\
S_{10}(t+1)&=& F_{10}(a(10,8)S_8(t)+a(10,11)S_{11}(t)-b(10,5)S_{5}(t)-b(10,7)S_{7}(t)-b(10,9)S_{9}(t))
\label{eq.five-b}.
\end{eqnarray}

In the paper of Li et al. \cite{li}, for all links one has $a(i,j)=a_g$ and $b(i,j)=a_r$, with $a_g=a_r=1$ 
for the majority of results.  Now we will study the general inhomogeneous case. Set new values to the 
coefficients means change the dynamical system definition. But will these value changes actually 
define new dynamical systems? Let $U$ be a subset of $\Omega$. If $F(U)=F'(U)$ for all subset $U$,  
the dynamical systems $(\Omega,F)$ and $(\Omega,F')$ are the same. In the inhomogeneous model,
 the dynamical system is defined by the set $C=\{K_{i,j}; 0 \leq i \leq 11, 0 \leq j \leq 11\}$. 
 Is that possible that a class $[C]$, which can contain more than one set of coefficients $C$, 
 defines the same dynamical system? If this happens  the non redundants inhomogeneities can be set 
 by choosing only one element of each class $[C]$.

In order to find this minimal set we must consider tow points: (i) the rule for time evolution takes 
into account only the sign of the sum $\sum_j K_{i,j} S_j(t)$, and (ii) the dynamical systems A and B are 
the same if, and only if, $\vec{S}_A(t+1)=\vec{S}_B(t+1)$ for all possible common initial condition 
$\vec{S}(t)$ in the state space.

We will illustrate how to find the minimal set explicitly for the dynamic of node $7$, which is in 
equation \ref{eq.trivial}. Note that both terms in the sum have the same sign. Whatever the values assumed by 
the pair $C=\{a(7,10),a(7,11)\}$, all possible initial state combinations $(S_{10}(t),S_{11}(t))$, 
namely $D=\{(0,0),(1,0),(0,1),(1,1)\}$, will result on the same case (positive, negative or zero) 
for the sum $\sum_j K_{i,j} S_j(t)$, namely $S=\{0,+,+,+\}$. But only the sign of the sum is important 
for the evaluation of $S_7(t+1)$, and we have that all the four possible values of the pair $C$ give 
the same map between $S$ and $D$. If two dynamical system have the same map between $S$ and $D$ they 
are identical. So we have all the four possible pairs defining the same dynamical systems what means 
that they take part of the same class $[C]$. We only need to take a representative pair of this class, 
being the simplest way to choose $C=(1,1)$. For all the other nodes the procedure is the same, but we 
use computer programs to make handle the calculations. The equation \ref{eq.trivial} have only one class,
 with all coefficients set equal one. The equations \ref{eq.two-a} and \ref{eq.two-b} have $3$ classes, 
 represented by $\{(1,1),(2,1),(1,2)\}$. The equation \ref{eq.three-a} and \ref{eq.three-b} have $11$ 
 classes, represented by
\[\{(1,1,1),(1,1,2),(1,1,3),(1,2,1),(1,2,2),(1,3,2),(2,1,1),(2,1,2),(2,2,1),(2,2,3),(3,1,2)\}.\]
The equation \ref{eq.four} has $67$ classes and the equations \ref{eq.five-a} and \ref{eq.five-b} 
have $3265$ classes. The minimal set that accounts for all the possible inhomogeneities has 
$3^2\cdot 11^2 \cdot 67 \cdot 3265^2= 7,77\times 10^{11}$ elements. This huge quantity means that we 
must use a numerical simulation in order to sample  this huge set.

\section{Results for the inhomogeneous model}

  First let us remind briefly the main results obtained by Li et al. \cite{li}
  for the homogeneous model with $t_d=1$.  From the $2,048$ initial conditions, $1,764$ converge to
  the fixed point $I_1=272$ representing the biological $G_1$ stationary state. The other initial conditions
  converge to six fixed points ($I_2=0$, $I_3=12$, $I_4=16$, $I_5=256$, $I_6=258$ and $I_7=274$).
  The pathway of the cell-cycle network is started by perturbing
  the $G_1$ stationary state: Cln3 is turned on.
 Then the state of the model returns to the $G_1$ fixed point after  $12$ time steps. The evolution
  of the proteins (or protein complexes), the so-called pathway of the cell-cycle network
  $Path_{homog}=[273, 278, 286, 14, 142, 1678, 1736, 1632, 1888, 1376, 368, 304, 272]$,
   follows the biological cell-cycle sequence.

We study the inhomogeneous model by randomly choosing the coefficients from the minimal set of interactions.
  In a typical numerical study we have $10^6$ samples.
  For each set of coefficients we study the evolution from each one of the 2,048 initial
  conditions and determine the fixed points (or the rare period-2 and period-3 cycles)
  with its attraction basin.
  The $G_1$ fixed point 272 is always present. In our simulations we find that its minimal  basin size
  is $BS_{min}(272)=11$. It occurs only in $0.34\%$ of the minimal interaction set.
  More rare, the maximal basin size $BS_{max}(272)=1999$ found in our simulations, appears
  in $0.23\%$ of the interactions cases. In this case, generally we have only 5 fixed points, instead of
  7 found in the homogeneous model.
  The mean basin size of the $G_1$ fixed point, $BS_{mean}(272)=1622.9$, is
  smaller than that of the Li et al. model ($BS_{homog}(272)=1764$). However, $65.3\%$ of
  the  interaction's minimal set have $BS(272) > 1764$.

We study also the pathway of the cell-cycle network. For each sample of coefficients, the temporal
  evolution of the pathway begins with the perturbed state 273, which is the stationary G1 state ($272$) 
  with the  Cln3 activation. Then we determine the pathway and its time step.
  In $10^6$  samples we find  336 different pathways. The
  pathway $Path_{freq}=[273, 278, 286, 14, 142, 1678, 1736, 1632, 1904, 1392, 368, 304, 272]$
   with 12 time steps is the most frequent. It appears in $10.9\%$ of the minimal interaction set,
  while the pathway of the Li et al.
  model appears only in $3.6\%$ of the cases. If we evaluate the average basin size of the
  $G_1$ fixed point in the ensemble of the  cases having the most frequent pathway, we
   find that $BS_{freq}(272)=1866.7$. Note that this average value is large than $BS_{homog}(272)=1764$.
   The only difference between $Path_{freq}$ and $Path_{homog}$ is that in the phase M the
   cyclin complex Cdh1 is turned on at time step 9 in $Path_{freq}$ instead of time step 11 in
   $Path_{homog}$, as it is shown in table \ref{orbit}. Is this difference meaningfull?

\begin{table}
\scalebox{0.8}{
\begin{tabular}{lccccccccccccc}
Time & Cln3 & Mbf & SBF & Cln1-2 & Cdh1 & Swi5 & Cdc20 & Clb5-6 & Sic1 & Clb1-2  & Mcm1 & Phase& I\\ \hline
1&1&0&0&0&1&0&0&0&1&0&0&START&273\\
2&0&1&1&0&1&0&0&0&1&0&0&G1&278\\
3&0&1&1&1&1&0&0&0&1&0&0&G1&346\\
4&0&1&1&1&0&0&0&0&0&0&0&G1&14\\
5&0&1&1&1&0&0&0&1&0&0&0&S&142\\
6&0&1&1&1&0&0&0&1&0&1&1&G2&1678\\
7&0&0&0&1&0&0&1&1&0&1&1&M&1736\\
8&0&0&0&0&0&1&1&0&0&1&1&M&1632\\
9&0&0&0&0&0/1&1&1&0&1&1&1&M&1888\\
10&0&0&0&0&0/1&1&1&0&1&0&1&M&1376\\
11&0&0&0&0&1&1&1&0&1&0&0&M&368\\
12&0&0&0&0&1&1&0&0&1&0&0&G1&304\\
13&0&0&0&0&1&0&0&0&1&0&0&STATIONARY PHASE &272\\ \hline
\end{tabular}}
\caption{Orbit with initial configuration being the perturbed stationary state. The column ``Time'' 
corresponds to the time steps; the columns identified by the nodes correspond to the time evolution 
of this nodes; the column ``Phase''  corresponds to the cell cycle phases; the column ``I'' 
correspond to the integer notation for the vector states. For Cdh1 in times $8$ and $9$ the 
notation $0 / 1$ means that the left state corresponds to $Path_{homog}$ and the right one to 
$Path_{freq}$.}\label{orbit}

\end{table}

The exit of phase M is controlled by the complex APC, which is activated mainly
   by Cdc20 and Cdh1. The protein Cdc20 actives APC that begins the degradation of Clb2 during
   the transition metaphase-anaphase. The second phase of the Clb2 degradation occurs by the
   linking of Cdh1 with APC, during the telophase. Therefore, the Clb2 degradation needs a
   sequential action \cite{lucas1}: (i) Cdc20 activating APC and (ii) Cdh1 linking to APC.
   Although the degradation by APC-Cdc20 is enough to the exit of phase M, the degradation
   by APC-Cdh1 is essential to keep the G1 phase stable \cite{lucas2}.
    In the homogeneous model Cdh1 is turned on simultaneously with the turning off of Clb2,
    implying that the second degradation mechanism is absent. On the other hand, this
    second mechanism is present in $Path_{freq}$ because Cdh1 and Clb2 are both turned on
    for two time steps.
   We can conclude that the early activation
   of Cdh1 is more coherent with the Clb2 degradation mechanism.

This results show that not just the big basin of attraction of the stationary state is indeed robust 
against changes in the coefficients, but  with inhomogeneities this basin of attraction gets 
bigger: $65,5\%$ of the interaction's minimal set have basin size bigger than $1764$, with a 
maximum value of $1999$. 

\section{Conclusions}
We  found that the big basin of attraction is indeed robust against change in the parameter $K_{i,j}$: 
$65,3\%$ of the  minimal interaction set have $BS(272)>1764$ and the maximum size being $1999$. 
Considering the orbit of the perturbed stationary state $273$ we found that the most frequent 
orbit ($P_{freq}$) is not the one exhibited by the homogeneous model: the orbit $P_{freq}$  
appears in $10,9\%$ of the minimum coefficient set while $P_{homog}$ appears only in $3,6\%$, 
that is, this orbit is more robust against changes on the coefficients. Besides that, the subset 
of the dynamical systems exhibiting $P_{freq}$ has a  mean basin size $BS_{freq}=1866,7$  that is 
bigger than $BS_{homog}=1764$. Not just more robust, but $P_{freq}$ has a better biological 
significance as we have shown by means of the earlier activation of Cdc1.

\bigskip

\noindent {\bf Acknowledgements}

\smallskip

\noindent {\small
The authors thank Alcides Castro e Silva for a careful reading of the manuscript.
JKLS thanks to  CNPq and FAPEMIG, Brazilian agencies. LW acknowledges CNPq
by financial support. }

\noindent {\small $^\ast$To whom correspondence should be addressed (e-mail: jaff@fisica.ufmg.br).}

\vfill
\end{document}